\documentclass[aip, apl, amsmath, amssymb, reprint]{revtex4-1}

\usepackage{graphicx}
\usepackage{hyperref}
\usepackage{dcolumn}
\usepackage{bm}

\usepackage[utf8]{inputenc}
\usepackage[T1]{fontenc}
\usepackage{mathptmx}
\usepackage{etoolbox}
\usepackage{xcolor}

\makeatletter
\def\@email#1#2{%
 \endgroup
 \patchcmd{\titleblock@produce}
  {\frontmatter@RRAPformat}
  {\frontmatter@RRAPformat{\produce@RRAP{*#1\href{mailto:#2}{#2}}}\frontmatter@RRAPformat}
  {}{}
}%
\usepackage{subfig}
\makeatother
\begin{document}

\preprint{AIP/123-QED}

\title{Kinetic Inductance and nonlinearity of MgB$_2$ Films at 4K}

\author{J. Greenfield}
 \email{jonathan.d.greenfield@jpl.nasa.gov.}
\affiliation{Jet Propulsion Laboratory, California Institute of Technology.}%
\affiliation{School of Earth and Space Exploration, Arizona State University.}%

\author{C. Bell}
\affiliation{Department of Physics, Arizona State University.}%

\author{F. Faramarzi}
\affiliation{Jet Propulsion Laboratory, California Institute of Technology.}%

\author{C. Kim}
\affiliation{Jet Propulsion Laboratory, California Institute of Technology.}%

\author{R. Basu Thakur}
\affiliation{Jet Propulsion Laboratory, California Institute of Technology.}%

\author{A. Wandui}
\affiliation{Jet Propulsion Laboratory, California Institute of Technology.}%

\author{C. Frez}
\affiliation{Jet Propulsion Laboratory, California Institute of Technology.}%

\author{P. Mauskopf}
\affiliation{School of Earth and Space Exploration, Arizona State University.}%
\affiliation{Department of Physics, Arizona State University.}%


\author{D. Cunnane}
\affiliation{Jet Propulsion Laboratory, California Institute of Technology.}%

\date{\today}

\begin{abstract}
We report on the fabrication and characterization of superconducting magnesium diboride (MgB$_2$) thin films intended for quantum-limited devices based non-linear kinetic inductance (NLKI) such as parametric amplifiers with either elevated operating temperatures or expanded frequency ranges. In order to characterize the MgB$_2$ material properties, we have fabricated coplanar waveguide (CPW) transmission lines and microwave resonators using $\approx$ 40 nm thick MgB$_2$ films with a measured kinetic inductance of $\sim$ 5.5 pH/$\square$ and internal quality factors,  $Q_i \approx 3 \times 10^4$ at 4.2 K. We measure the NLKI in MgB$_2$ by applying a DC bias to a 6 cm long by 4 µm wide CPW transmission line, and measuring the resulting phase delay caused by the current dependent NLKI. We also measure the current dependent NLKI through CPW resonators that shift down in frequency with increased power applied through the CPW feedline. Using these measurements, we calculate the characteristic non-linear current parameter, $I_*$, for multiple CPW geometries. We find values for corresponding current density, $J_* = 12-22$~MA/cm$^2$ and a ratio of the critical current to the non-linear current parameter, $I_C/I_* = 0.14-0.26$, similar to or higher than values for other superconductors such as NbTiN and TiN.  
\end{abstract}

\maketitle

\begin{figure}[ht]
    \centering
    \subfloat{\includegraphics[width=0.9\columnwidth]{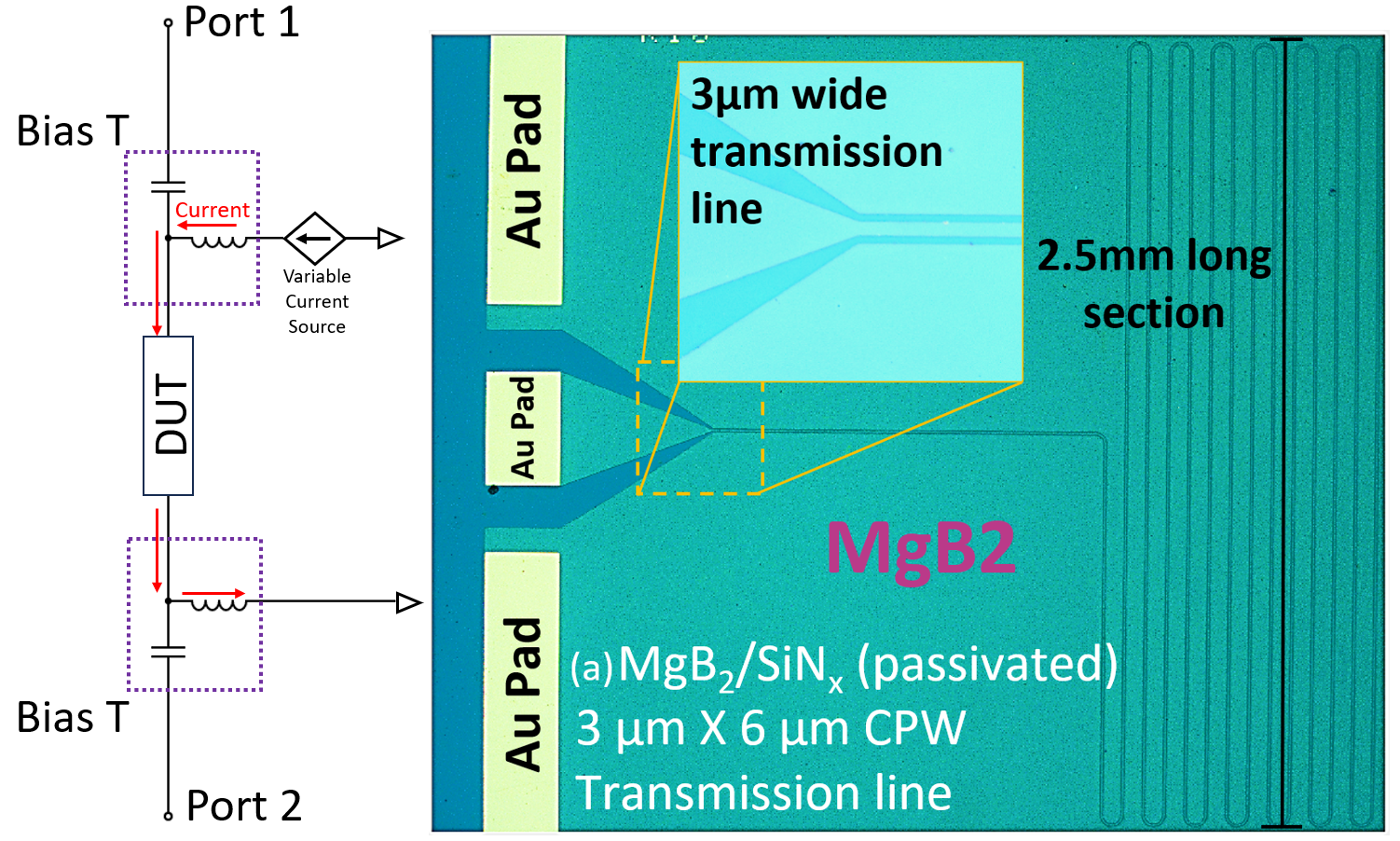}}
    \caption{Schematic of our test setup for DC biasing a CPW transmission line. The optical image on the right is similar to our CPW transmission line. The measured devices were widened to 4 $\mu$m to improve impedance matching. Each meander section is roughly 2 mm long, giving the transmission line a total length of 6 cm.}    
    \label{fig:fabpics}
\end{figure}

\noindent Kinetic inductance in thin-film superconductors has enabled a wide class of ultra-sensitive photon and phonon detectors in the form of resonators\cite{day2003broadband, jonasannual, sergeev2002, Doyle2008, perido2024parallel, mazin2009mkid, vissers2020ultrastable}. More recently, coherent devices such as parametric amplifiers, frequency converters, and field sensors have been developed, making use of the current-driven non-linearity of the kinetic inductance, similar to the nonlinear inductance in Josephson Junctions\cite{ho2012,4-8fara,shu2021nonlinearity,malnou2021three,Cunnane2024, KPUP,sypkensKIM,shu2021nonlinearity,malnou2021three,4-8fara,faramarzi2024nearquantumlimitedsubghz,JPA,JPA2}. This enables similar performance to devices using Josephson Junctions with different fabrication and optimization constraints, particularly related to new materials.
\newline
\newline
Near quantum-limited kinetic inductance traveling-wave parametric amplifiers (KI-TWPA) with a bandwidth of a few GHz\cite{ho2012,4-8fara,shu2021nonlinearity,malnou2021three}, frequency multipliers that enable high efficiency harmonic generation\cite {Cunnane2024}, and even current sensing devices similar to RF SQUID readouts \cite{KPUP,sypkensKIM} have all been demonstrated using non-linear kinetic inductance (NLKI). As of yet, these devices have been fabricated using NbTiN \cite{shu2021nonlinearity,malnou2021three,4-8fara}, NbN\cite{sypkensKIM}, or TiN \cite{faramarzi2024nearquantumlimitedsubghz} as the high kinetic inductance material. Most of these materials have inherent kinetic inductance around 5-40 pH/$\square$ for a film thickness between 10-40 nm. In this work, we show that similar values can be achieved in MgB$_2$ despite the factor of 2-3 higher critical temperature than NbN or NbTiN, enabling access to higher frequency THz ranges or higher operating temperatures. The MgB$_2$ thin films used for this work have a $T_c$ of 29 K, compared to 14 K for NbTiN used in similar work. We describe the fabrication process, microwave quality factors, and non-linearity of these films through characterization of coplanar waveguide (CPW) transmission lines and quarter-wave resonators. 
\newline
\newline
The kinetic inductance of a transmission line depends on the bias current $I$ and nonlinearity scale-factors $I_*$ and $I_{*}'$ (Eq.~\ref{eqn:eq1}). Material properties and device geometry determine these characteristic currents \cite{shu2021nonlinearity}.
\begin{equation}
\label{eqn:eq1}
    \displaystyle{\mathcal{L}(I) = \mathcal{L}_0\left[1 + \left (\frac{I}{I_*}\right)^2 + \left(\frac{I}{I_*^{\prime}} \right)^4 + \ldots \right]}
\end{equation}
Introducing a current bias leads to inductance-induced phase delay, $\phi (I) = 2\pi f\ell \sqrt{\mathcal{L}(I)\mathcal{C}}$ through a transmission line of length $\ell$; here, $\mathcal{L}$ and $\mathcal{C}$ are inductance and capacitance per unit length of the transmission line and $f$ is the signal frequency.  
\newline
\newline
Previous work on these MgB$_2$ films\cite{MgB2paper} has demonstrated devices operating at 4.2 K and shown the fabrication maturity needed to realize practical devices.
MgB$_2$ has two distinct energy gaps associated with different bands: a lower-energy pi ($\pi$) band,  $\Delta_\pi = 2.1$ meV, and a higher-energy sigma ($\sigma$) band, $\Delta_\sigma = 7.0$ meV, leading to a relatively high critical temperature ($T_c$). It has been shown that in the limit that the electron mean free path ($l$) is much shorter than the coherence length ($\xi$), the interaction between the two gaps leads to anomalous behavior in MgB$_2$, with gap energy falling somewhere between the two individual gaps \cite{sym11081012}. Projecting the superconducting gap properties of MgB$_2$ could facilitate the development of novel devices capable of operating beyond 1 THz, advancing sub-millimeter astronomy.
\newline
\newline
In order to characterize the superconducting properties of MgB$_2$ films and transmission lines, we fabricate coplanar waveguide devices from uniform thin films. We start by co-sputtering Mg and B followed by an in situ B capping layer. The Mg/B layer is deposited to be slightly sub-stoichiometric to achieve a kinetic inductance similar to NbTiN \cite{shu2021nonlinearity}. We then use a rapid thermal process (RTP) to react the constituent elements into MgB$_2$. The Boron capping layer is easily removed in Cl$_2$ plasma, followed by the deposition of a PECVD SiN$_x$ hard mask layer. The CPW device patterns were lithographically defined using a maskless lithography tool and then transferred into the SiN$_x$ layer using a Fluorine based RIE etch. Once the pattern is transferred into the SiN$_x$, we use a BCl$_3$ plasma etch, typically used to etch Al, to transfer the pattern to the MgB$_2$. We then passivate the devices with approximately 200 nm more of PECVD SiN$_x$ to ensure sidewall coverage. A second lithography step is carried out to negatively define the Au bond pads. A second Fluorine etch removes the Nitride and an in situ ion beam etch is carried out prior to the deposition of the Au pads to ensure good Ohmic contact. Fig.~\ref{fig:fabpics} shows an image of a 6 cm-long transmission line (TL) meandered across the 3 mm x 5 mm chip. The CPW dimensions are 4 µm center conductor with 2 µm gaps to the ground plane on either side. 
\newline
Previous devices were fabricated using a different method  \cite{MgB2paper}, using a Ti/Au hard mask and an ion mill for etching the pattern into the MgB$_2$. Films fabricated using this method produced resonators with internal quality factors (Q$_i$) as high as 10,000 at 4.2 K \cite{MgB2paper}. However, we have since developed a new technique that eliminates the use of the ion mill and more consistently yields resonators with higher internal quality factors, even above 30,000 at 4.2 K. In this work, we present the results of those fabrication improvements and take the first step toward practical MgB$_2$ devices by demonstrating a similar level of non-linearity as those reported in the literature for more mature superconducting materials and devices.
\vspace{-4mm}
\newline
\newline
\noindent
\begin{figure}[h!]
    \centering
    \subfloat{\includegraphics[width=0.99\columnwidth]{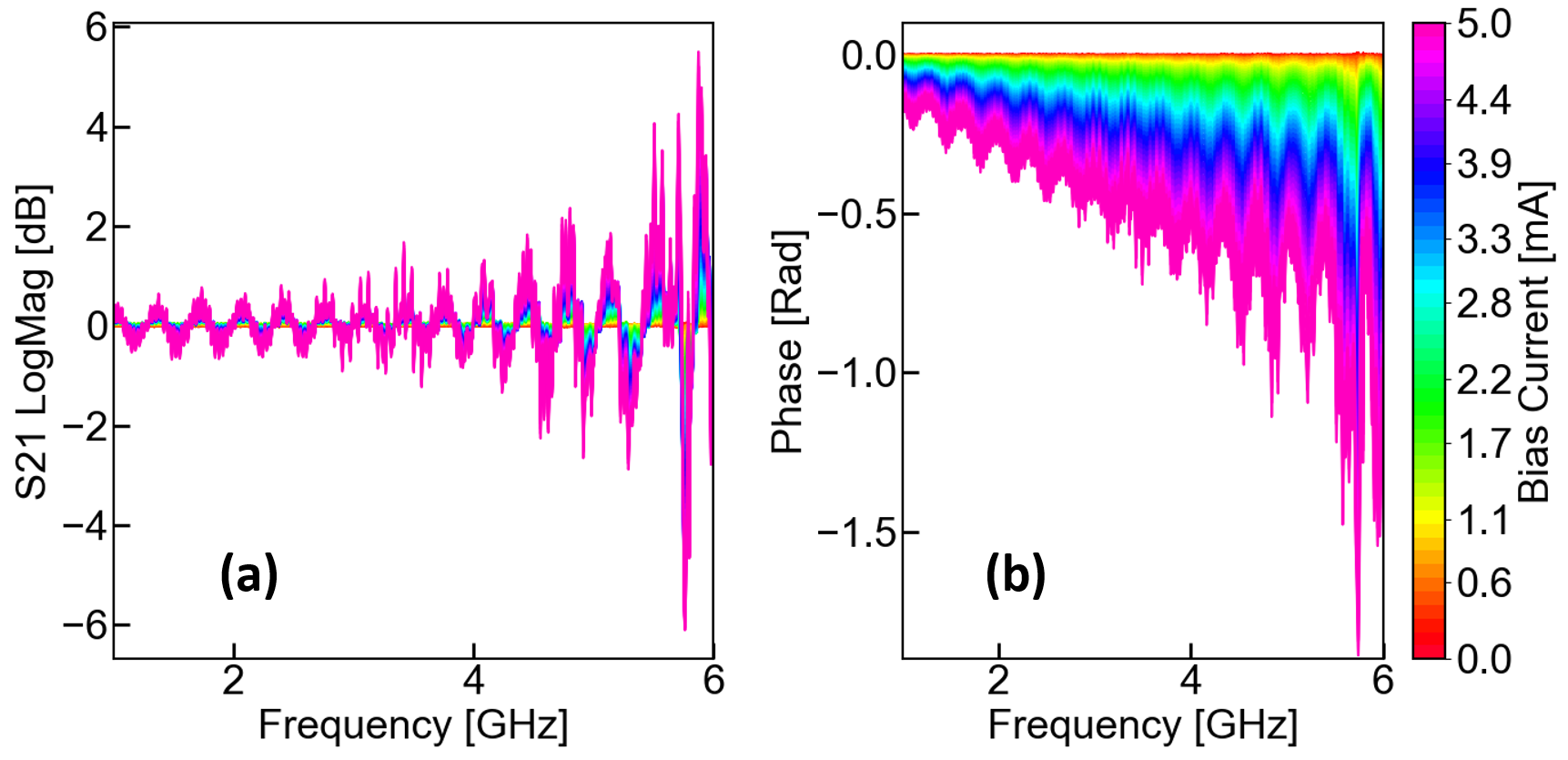}}
    \hfil
    \subfloat{\includegraphics[width=0.99\columnwidth]{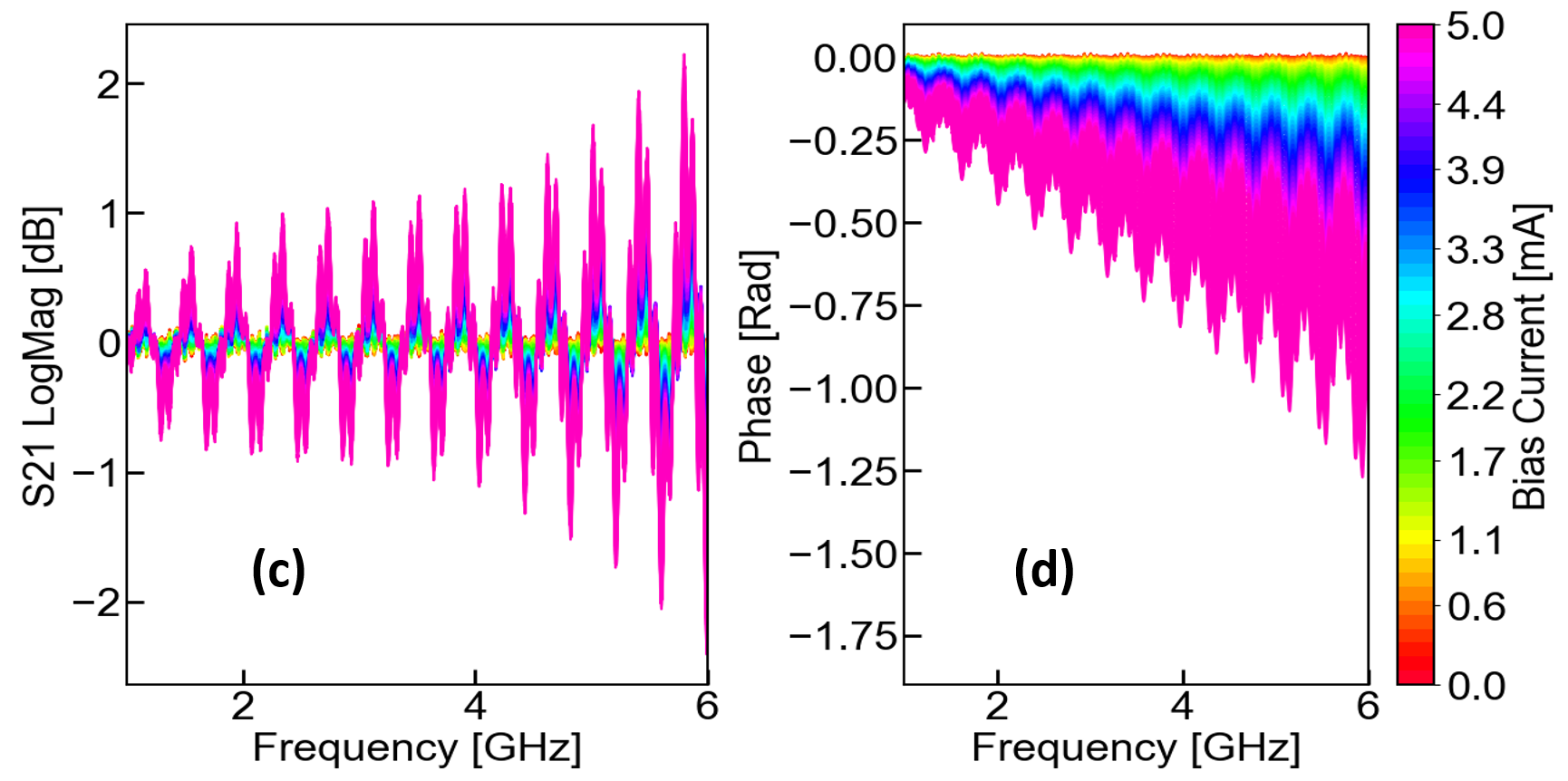}}
    
    \caption{ S21 Log magnitude transmission (a) and phase (b) as a function of frequency for an MgB2 CPW transmission line. The data is normalized to the zero current data, and color scale shows the bias current in the transmission line. (c \& d) Simulation of the same system using Scikit RF \cite{9632487} The simulation includes The coax cables and PCB on either side of the DUT as well as the room temperature amplifier directly before the input of the VNA.}    
    \label{fig:fabpic_b} 
\end{figure}

\begin{figure}[h]
    \centering
    \subfloat{\includegraphics[width=\columnwidth]{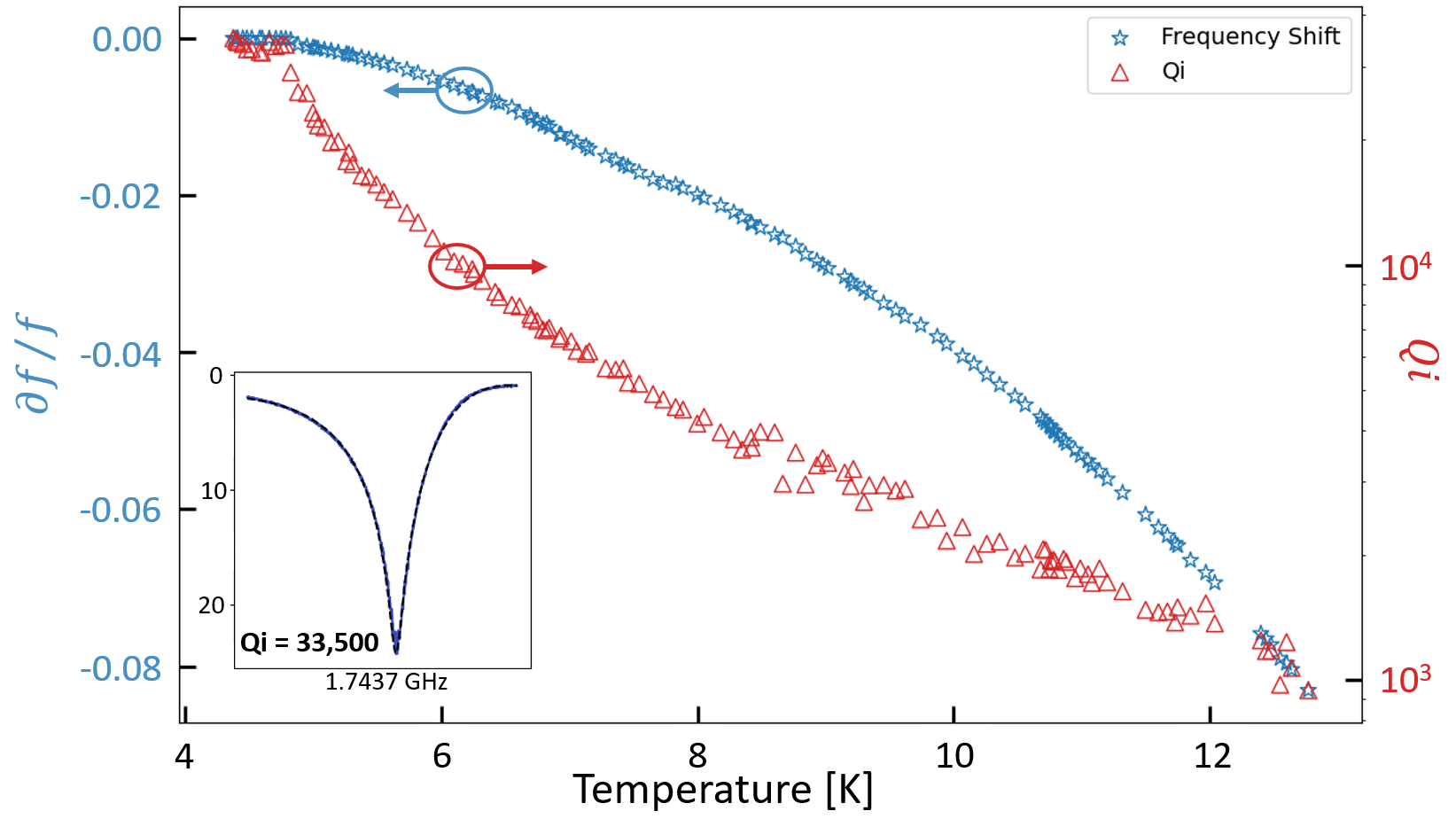}}
    \hfil
    \caption{The left y-axis (blue stars) is the fractional frequency shift versus temp of a MgB$_2$ lumped element resonator from 4.2 K to 13 K. The inductor is 3 $\mu$m wide, and the capacitance comes from an IDC. The right y-axis (red triangles) is the resonator's internal quality factor Q$_i$ as a function of temperature. Inset is a $S_{21}$ graph highlighting the resonator's Q$_i$ $\sim 3.3 \times 10^4$ at 4.3~K, where the blue line represents the experimental data, and the dashed line indicates the fit using standard fitting formulation \cite{Carter2016}.}
    \label{fig:tempdep}
\end{figure}

\noindent To characterize and optimize losses in this material system, we have fabricated both distributed quarter-wave (QW) CPW TL resonators as well as lumped-element (LE) resonators coupled to a CPW feedline. In the QW resonators, we have achieved internal quality factors around Q$_i$ $\sim 1.3 \times 10^4$, largely independent of the CPW geometry. This limit is likely set by the interface between the superconductor and the substrate, which can be more lossy than other materials due to the high temperature annealing of the films. The nature of the distributed resonator is such that current crowding is likely to occur near the shorted end of the resonator while the field is highest/densest near the open end of the resonator. In our QW resonators, each open end is coupled to the feedline with an identical coupling inductor that is  capacitively coupled. The fact that we do not see different losses from one resonator geometry to the next implies that the loss could indeed be tied to the coupling. Using a lumped-element architecture, we can improve on this because an inter-digitated capacitor (IDC) with wide spacing can reduce current crowding in a specific location, reducing field density in the lossy interface.  In contrast, LE resonators empirically show higher internal quality factors, as seen in Fig.~\ref{fig:tempdep}, so we use them to report on the highest values currently achieved. 
\newline
\newline
\noindent
The resonator's Q$_i$ includes all loss mechanisms relevant to a transmission line of the same geometry. We expect improvements on these quality factors; however, we consider how the loss will impact transmission on a long line, taking $\mathrm{tan}(\delta) = 1/Q_i$. We then use the approximation for transmission line loss, $RL = 27.3\times \sqrt{\epsilon_R}\times f\times \mathrm{tan}(\delta) /{c}$ [dB/m]
\cite{Pozar:882338}. Using Q$_i$ $\sim 1.3 \times 10^4$, this gives 0.15 dB/m in a CPW transmission line, measured for lines up to 4 GHz. For a practical device such as a variable 2$\pi$-phase shifter based on the parameters obtained in this work, a CPW TL would need to be 22 cm long, and would only give 0.03 dB of insertion loss.
\newline
\newline
\noindent
We measured the scattering parameters as a function of frequency and DC bias current in a 40 nm thick MgB$_2$ CPW meandering transmission line (shown in Fig 1). The CPW meander has a center conductor width of ~4 µm, a gap of ~2 µm, and a length of $\ell = 64.6$~mm. From this geometry we obtain a propagation speed, $c_\phi = 1/\sqrt{\mathcal{L}\mathcal{C}}$, of 0.19$c$. The kinetic inductance fraction, $\alpha$, is measured to be 0.76 from resonators fabricated in the same batch. The CPW devices were measured in a liquid helium Dewar at 4.2 K using a Vector Network Analyzer (VNA) with resolution of 100 KHz from 1-6 GHz. The transmission line was DC-biased across a 1 k$\Omega$ resistor and a voltage source. For $I^2/I_*^2 \ll 1$, the phase shift is given by:
\newline
\begin{equation}
    \frac{\delta \phi}{\nu} = - \frac{\pi \ell}{c_\phi}\frac{\delta \mathcal{L}}{\mathcal{L}_0} \approx - \frac{\alpha \pi \ell}{c_\phi}\frac{I^2}{I^{*2}}
\end{equation}
\newline
where $\nu$ is the measurement frequency, $\mathcal{L}_0$ is the inductance per unit length at zero bias and $\delta \mathcal{L}$ is the change in kinetic inductance due to the DC current. We then plot the phase change over the applied current and fit the curve to the modified equation to extract $I_*$.
\newline

\begin{figure}[ht]
    \centering
    \subfloat{\includegraphics[width=0.99\columnwidth]{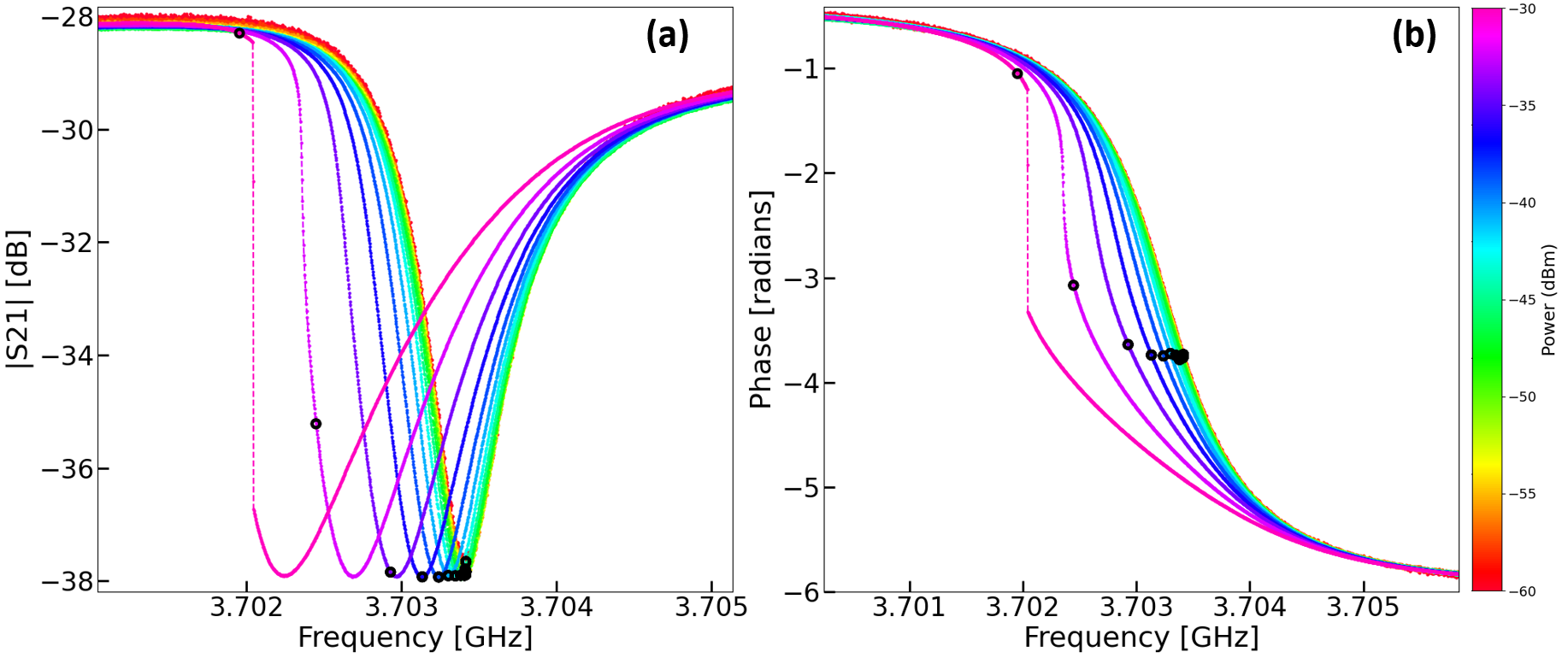}}
    \caption{(a) Plot of S$_{21}$ (dB) vs. frequency (GHz) of a 5$\mu$m wide distributed resonator as a function of current. The black dots on the graph represent the location of the resonant frequency of the resonator. (b) the unwrapped phase (Rad) of the biased resonators and the black dots also represent the location of the resonant frequency of the resonator. The different colors indicate a different current bias in the resonator.}
    \label{fig:phase_shift}
\end{figure}

\noindent
The variations in the measured magnitude and phase as a function of frequency are primarily due to standing waves, which result from an impedance mismatches caused by the high kinetic inductance in the material. 
Impedance matching of these MgB$_2$ lines within the CPW architecture poses challenges, largely due to significant kinetic inductance. Future designs will address this challenge by incorporating capacitive fingers along the transmission line to more effectively achieve a characteristic impedance of 50 $\Omega$. Assuming a 1 um linewidth, this will increase the kinetic inductance by a factor of 4, achieve near unity kinetic inductance participation, and slow the propogation speed by a factor of 8, 
\newline
\newline
\noindent
To add to the fidelity of the transmission line measurements, we measure the non-linearity in our MgB$_2$ films as a function of geometry using a set of QW CPW resonators with varying conductor widths to compare values for $I_*$. Each chip was fabricated with five resonators, featuring an identical coupling inductor tapered to center conductor widths of $w = 3$~µm, 4~µm, 5~µm, 6~µm, and 10~µm respectively. All inductors on these resonators were designed with a uniform length $\ell = 4.175$~mm, a fundamental resonant frequency $f_{0}^{geom} =  7.039$~GHz, and an impedance $Z_0 \approx$ 50 Ohms. The gap spacing $s$ was designed so that the zero-kinetic inductance resonant frequency and impedance was independent of $w$, and that the corresponding capacitance per unit length and geometric inductance per unit length were $\mathcal{C} = 1/(4 f_0 Z_0 \ell) = 170$~pF/m and $\mathcal{L}_G$ = 428~nH/m, respectively.
\newline
\newline
\noindent
We used a liquid Helium dip probe and calibrated the input power P$_{in}$ at the device using a microwave power meter. Following the calibration, the resonators were cooled to 4.2 K, at which point we biased the 10 µm wide transmission line until it transitioned to a normal state, corresponding to a critical current $I_c = 12$ mA $\pm 0.12$ mA , or $J_c = 3$ MA/cm$^2$. We confirmed our power calibration to the chip by driving the feedline normal with RF current. Subsequently, we conducted power sweeps across all five of the resonators on the chip, ranging from -70 dBm to -40 dBm in 2 dBm increments. From the collected data, we extracted key parameters, such as the superconducting resonant frequency $f_{0}$, the internal quality factor Q$_i$, and the coupling quality factor Q$_c$, as functions of input power. We employed a fitting routine based on the method developed by Swenson et al.\cite{swenson2013} to accurately determine the relevant data and extract the nonlinearity parameter, (a). The quality factors were derived by analyzing the geometry of the I-Q circle\cite{Carter2016,probst2014}. As demonstrated in Fig.~\ref{fig:phase_shift}a \& b, this approach successfully fits resonators with nonlinearity parameters up to a = 1.3\cite{swenson2013}. 
\newline

\begin{figure}[ht]
    \centering
    \subfloat{\includegraphics[width=0.99\columnwidth]{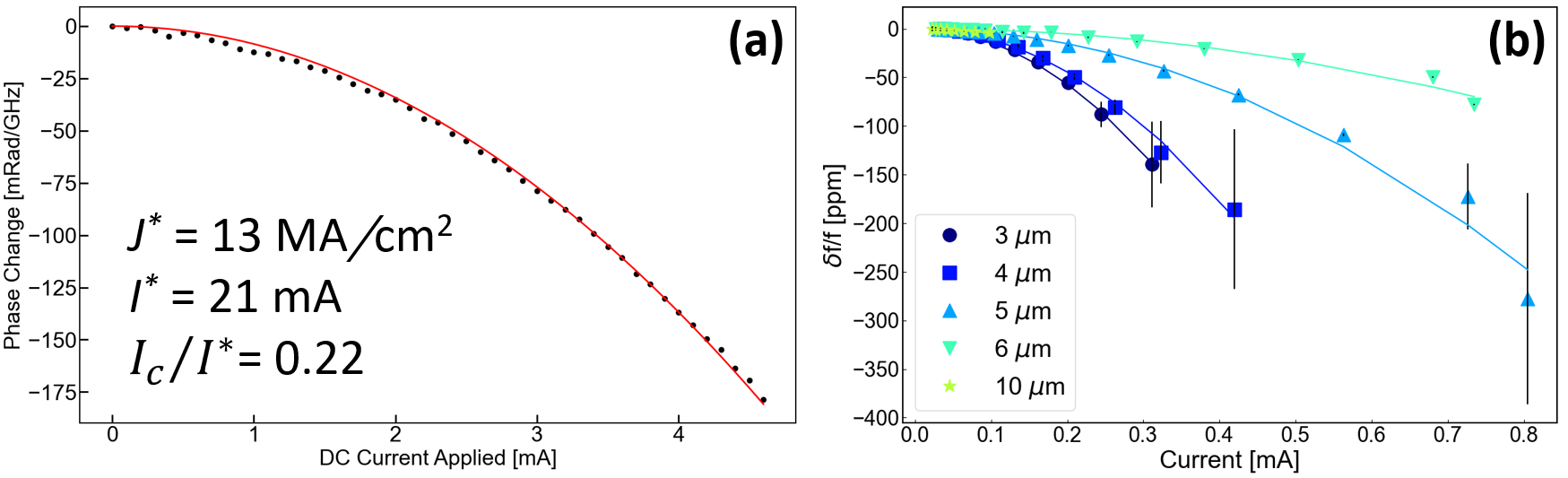}}
    \caption{(a) Plot of the phase change (mRad/GHz) vs. applied DC current (mA) fitted to extract $I_*$. (b) Plot of the relative frequency shift for resonators of different widths vs. estimated current in the resonator, fitted to extract $I_*$. The error bars reflect the best fit uncertainities in the bifurcation parameter from fitting the resonator using the Swenson model\cite{swenson2013}.}
    \label{fig:shift}
\end{figure}

\noindent
 From the measured resonant frequencies of the different geometries at low input power, we determined the kinetic inductance fraction $\alpha = 1 - (f_0/f_0^{geom})^2$ for each CPW resonator. Using these values for $\alpha$, we calculated the kinetic inductance per square. These inductance values allowed us to compute the total inductance $L_{Total}$ and the corresponding impedance Z$_L$ for each resonator on the chip. Consequently, we derive the effective penetration depth \cite{day2003broadband} $L_k \sim \mu_0 \lambda_{eff}$. In the case where the film thickness is much less than the penetration depth, we can approximate that $\lambda_{eff} = \lambda^2/d$, where $d$ is the thickness of the material. Finally, we calculate the penetration depth as $\lambda = \sqrt{(L_kd)/\mu_0}$, which gives us an average $\lambda \approx$ 420 nm.
\newline
\newline
\noindent
Similar to previous reports for NbTiN\cite{dai2023method}, we calculated the current stored in the resonator on resonance as a function of input power ($P_{in}$), $I^2$. For a distributed resonator, the equivalent lumped capacitance is half of the value calculated from the capacitance per unit length due to the non-uniform field distribution. Therefore:
\begin{equation}
    I^2 = \frac{\omega_r \mathcal{C} \ell P_{in}Q_l^2}{Q_c}
\end{equation}\
We then determine $I_*$ using the following equation: 
\begin{equation}
    \frac{\delta f}{f} (I(P_{in})) \approx -\frac{\alpha}{2} \frac{I(P_{in})^2}{I_*^2}
\end{equation}
Results from these measurements are given in Table~\ref{tab:FittingStats}. The primary source of error in our calculations stems from power calibrations. The same power range was used for all resonators, leading to very small shift in the 10 um resonator, which made a fitting to extract $I_*$ unreliable. 
\newline
\begin{table}[ht]
\caption{Measured properties of CPW resonators (R) and transmission lines (TL).}

\begin{center}
\begin{tabular}{|l|c|c|c|c|c|c|} 
 \hline
 R/TL & Width/Gap & $\alpha$ & $f_0$ & $L_k^{\text{sq}}$ & $I_*$ & $I_c/I_*$  \\
 & ($\mu$m) & & (GHz) & (pH/sq) & (mA) &  \\
 \hline
R & 3/2  & 0.81 & 3.083 & $5.47 \pm 0.2$ & $17 \pm 3.3$ & 0.211 \\
R & 4/2.7  & 0.76 & 3.433 & $5.42 \pm 0.3$ & $19 \pm 3.6$ & 0.253 \\
R & 5/3.3  & 0.72 & 3.703 & $5.50 \pm 0.2$ & $31 \pm 6.3$ & 0.194 \\
R & 6/4  & 0.69 & 3.930 & $5.71 \pm 0.3$ & $52 \pm 7.8$ & 0.139 \\
TL & 4/2  & 0.76 & - & $5.42 \pm 0.3$ & $21 \pm 1.3$ & 0.220 \\
 \hline
\end{tabular}
\label{tab:FittingStats}
\end{center}
\end{table}

\begin{figure}[ht]
    \centering
    \includegraphics[width=0.5\textwidth]{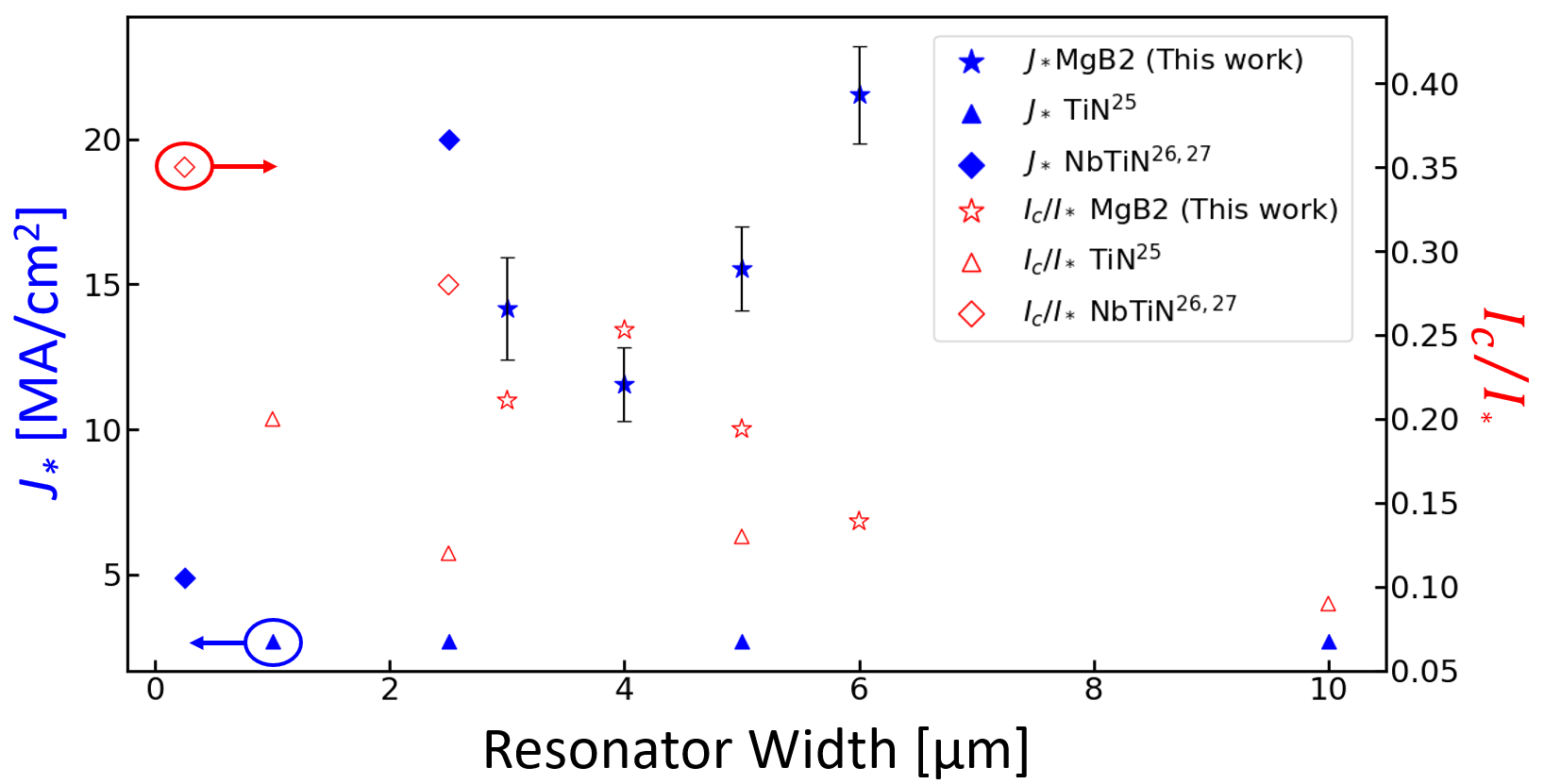}
    \caption{$J_*$ (left) vs width of resonator for MgB$_2$, TiN \cite{dai2023method}, \& NbTiN \cite{vissers2015,basu_thakur2019superconducting}. The right y-axis shows the corresponding $I_c$/$I_*$ ratio. This ratio is used as a metric to compare materials regardless of superconducting properties like $T_c$ or $L_k$. Our work ($\star$) is the only work reported at 4 K while all other work was performed at temperatures less than 1 K.}
    \label{fig:cpw_ms_comparison_new}
\end{figure}

\noindent
In Fig.~\ref{fig:cpw_ms_comparison_new}, we plot the results from our measurements of MgB$_2$ devices in comparison with properties of other superconducting materials. MgB$_2$ shows consistent characteristic current densities $J_*$ across the range of resonator widths ranging from 12-22 MA/cm$^2$. We also plot the ratio of $I_c/I_*$ for MgB$_2$ as a function of linewidth and see no significant trend. There are a number of factors that could impact this ratio, such as defects along the transmission line or non-uniform current densities. The Pearl length, $\Lambda = 2\lambda^2/d$, should be $\sim$ 8.4 $\mu$m, based on the calculated penetration depth, implying that current distribution should be fairly uniform in all but the 10 $\mu$m wide line.
 \newline
\newline
 \noindent
 We have characterized the NLKI in MgB$_2$ films using CPW transmission lines QW resonators. The scale of non-linearity is described by the expression from Eq.~\ref{eqn:eq1}. For comparison, the NbTiN microstrip devices developed by JPL, which are optimized for NLKI, achieve a non-linearity of 20$\%$ at the critical current \cite{MgB2paper}. Our results show MgB$_2$ devices have already achieved $\sim$5$\%$ on par with many recent reports\cite{dai2023method} and has the potential to increase with
 improved designs and fabrication process optimization. In addition, these MgB$_2$ thin films have a $T_c$ of approximately 29 K, compared to NbTiN’s $T_c$ of 14 K, making them suitable for applications that require higher operating temperatures or frequencies, such as those in the THz range. While we have observed $L_k$ values of about 5.5 pH/$\square$ for our 40 nm films, further adjustments in film stoichiometry could increase $L_k$ significantly, surpassing the kinetic inductance values achieved by NbTiN films of comparable thickness, while still having higher $T_c$.
\newline
\newline
\noindent
The measured NLKI has enabled the development of compact cryogenic phase shifters. We have demonstrated current-tunable phase shifts ranging from 0 to 0.75 radians per GHz and 4.2 K for CPW transmission lines. Longer or narrower transmission lines capable of achieving a $2\pi$ phase shift are anticipated to maintain negligible losses ($<.05$ dB). This conclusion is supported by the high-quality factor measurements of resonators fabricated similarly to the transmission lines and evaluated under the same conditions.
\newline
\newline
\noindent
We are constantly improving the quality factor that can be achieved through fabrication process improvements. Currently, the limitations on the quality factor are expected to be in the interface layer between the substrate stack and MgB$_2$ due in part to the high temperature post annealing process. Ongoing efforts are directed towards improving our fabrication processes to increase the ratio of $I_C/I_*$ and enabling longer lines with NLKI necessary for advanced applications such as kinetic inductance traveling-wave parametric amplifiers (KI-TWPAs), Superconducting On-chip Fourier Transform Spectrometers (SOFTS), and frequency translators. 
 \newline
\newline
 \noindent
This research was carried out at the Jet Propulsion Laboratory, California Institute of Technology, under a contract with the National Aeronautics and Space Administration (80NM0018D004). The work in this publication was supported by NASA’s Astrophysics Research and Analysis Program (APRA) under task order No. 80NM0018F0610, and proposal number 20-APRA20-0127. This work was made possible by the capabilities developed by Dr.\ Daniel Cunnane under a Nancy Grace Roman Early Career Technology Fellowship (RTF). Dr.\ Faramarzi's research was supported by appointment to the NASA Postdoctoral Program at the Jet Propulsion Laboratory, administered by Oak Ridge Associated Universities under contract with NASA. We acknowledge the support and infrastructure provided for this work by the Microdevices Laboratory at JPL, as well as the funding provided by the JPL SIP. We thank Rick LeDuc, Bruce Bumble, and Andrew Beyer for advice on device fabrication. © 2024. All rights reserved.
\bibliographystyle{ieeetr}
\bibliography{refs}

\begin{thebibliography}{10}

\bibitem{day2003broadband}
P.~K. Day, H.~G. LeDuc, B.~A. Mazin, A.~Vayonakis, and J.~Zmuidzinas, ``A
  broadband superconducting detector suitable for use in large arrays,'' {\em
  Nature}, vol.~425, no.~6960, pp.~817--821, 2003.

\bibitem{jonasannual}
J.~Zmuidzinas, ``Superconducting microresonators: Physics and applications,''
  {\em Annual Review of Condensed Matter Physics}, vol.~3, no.~Volume 3, 2012,
  pp.~169--214, 2012.

\bibitem{sergeev2002}
A.~Sergeev, V.~Mitin, and B.~Karasik, ``Ultrasensitive hot-electron
  kinetic-inductance detectors operating well below the superconducting
  transition,'' {\em Applied physics letters}, vol.~80, no.~5, pp.~817--819,
  2002.

\bibitem{Doyle2008}
S.~Doyle, P.~Mauskopf, J.~Naylon, A.~Porch, and C.~Duncombe, ``Lumped element
  kinetic inductance detectors,'' {\em Journal of Low Temperature Physics},
  vol.~151, pp.~530--536, Apr 2008.

\bibitem{perido2024parallel}
J.~Perido, P.~Day, A.~Beyer, N.~F. Cothard, S.~Hailey-Dunsheath, H.~Leduc,
  B.~H. Eom, and J.~Glen, ``Parallel plate capacitor aluminum kids for future
  far-infrared space-based observatories,'' {\em Journal of Low Temperature
  Physics}, 2024.

\bibitem{mazin2009mkid}
B.~A. Mazin, ``Microwave kinetic inductance detectors: The first decade,'' in
  {\em AIP Conference Proceedings}, vol.~1185, pp.~135--142, American Institute
  of Physics, 2009.

\bibitem{vissers2020ultrastable}
M.~R. Vissers, J.~E. Austermann, M.~Malnou, C.~M. McKenney, B.~Dober,
  J.~Hubmayr, G.~C. Hilton, J.~N. Ullom, and J.~Gao, ``Ultrastable
  millimeter-wave kinetic inductance detectors,'' {\em Applied Physics
  Letters}, vol.~116, p.~032601, 2020.

\bibitem{ho2012}
B.~Ho~Eom, P.~K. Day, H.~G. LeDuc, and J.~Zmuidzinas, ``A wideband, low-noise
  superconducting amplifier with high dynamic range,'' {\em Nature Physics},
  vol.~8, no.~8, pp.~623--627, 2012.

\bibitem{4-8fara}
F.~Faramarzi, R.~Stephenson, S.~Sypkens, B.~H. Eom, H.~LeDuc, and P.~Day, ``{A
  4–8 GHz kinetic inductance traveling-wave parametric amplifier using
  four-wave mixing with near quantum-limited noise performance},'' {\em APL
  Quantum}, vol.~1, p.~036107, 07 2024.

\bibitem{shu2021nonlinearity}
S.~Shu, N.~Klimovich, B.~H. Eom, A.~Beyer, R.~B. Thakur, H.~Leduc, and P.~Day,
  ``Nonlinearity and wide-band parametric amplification in a (nb, ti) n
  microstrip transmission line,'' {\em Physical Review Research}, vol.~3,
  no.~2, p.~023184, 2021.

\bibitem{malnou2021three}
M.~Malnou, M.~Vissers, J.~Wheeler, J.~Aumentado, J.~Hubmayr, J.~Ullom, and
  J.~Gao, ``Three-wave mixing kinetic inductance traveling-wave amplifier with
  near-quantum-limited noise performance,'' {\em PRX Quantum}, vol.~2, no.~1,
  p.~010302, 2021.

\bibitem{Cunnane2024}
D.~Cunnane, H.~Leduc, N.~Klimovich, F.~Faramarzi, A.~Beyer, and P.~Day,
  ``High-efficiency ka-band frequency multiplier based on the non-linear
  kinetic inductance in a superconducting microstrip,'' {\em Applied physics
  letters}, vol.~124, no.~0, p.~000000, 2024.

\bibitem{KPUP}
A.~{Kher}, P.~K. {Day}, B.~H. {Eom}, J.~{Zmuidzinas}, and H.~G. {Leduc},
  ``{Kinetic Inductance Parametric Up-Converter},'' {\em Journal of Low
  Temperature Physics}, vol.~184, pp.~480--485, July 2016.

\bibitem{sypkensKIM}
S.~Sypkens, F.~Faramarzi, M.~Colangelo, A.~Sinclair, R.~Stephenson, J.~Glasby,
  P.~Day, K.~Berggren, and P.~Mauskopf, ``Development of an array of kinetic
  inductance magnetometers (kims),'' {\em IEEE Transactions on Applied
  Superconductivity}, vol.~31, no.~5, pp.~1--4, 2021.

\bibitem{faramarzi2024nearquantumlimitedsubghz}
F.~Faramarzi, S.~Sypkens, R.~Stephenson, B.~H. Eom, H.~Leduc, S.~Chaudhuri, and
  P.~Day, ``A near quantum limited sub-ghz tin kinetic inductance traveling
  wave parametric amplifier operating in a frequency translating mode,'' 2024.

\bibitem{JPA}
J.~Aumentado, ``Superconducting parametric amplifiers: The state of the art in
  josephson parametric amplifiers,'' {\em IEEE Microwave magazine}, vol.~21,
  no.~8, pp.~45--59, 2020.

\bibitem{JPA2}
C.~Macklin, K.~O’brien, D.~Hover, M.~Schwartz, V.~Bolkhovsky, X.~Zhang,
  W.~Oliver, and I.~Siddiqi, ``A near--quantum-limited josephson traveling-wave
  parametric amplifier,'' {\em Science}, vol.~350, no.~6258, pp.~307--310,
  2015.

\bibitem{MgB2paper}
C.~Kim, C.~Bell, J.~Evans, J.~Greenfield, N.~Lewis, and D.~Cunnane,
  ``Wafer-scale magnesium diboride thin films with tunable high kinetic
  inductance,'' {\em ACS Nano}, vol.~18, no.~9, pp.~9625--9635, 2024.

\bibitem{sym11081012}
H.~Kim, K.~Cho, M.~A. Tanatar, V.~Taufour, S.~K. Kim, S.~L. Bud’ko, P.~C.
  Canfield, V.~G. Kogan, and R.~Prozorov, ``Self-consistent two-gap description
  of mgb2 superconductor,'' {\em Symmetry}, vol.~11, no.~8, 2019.

\bibitem{9632487}
A.~Arsenovic, J.~Hillairet, J.~Anderson, H.~Forst?n, V.~Rie?, M.~Eller,
  N.~Sauber, R.~Weikle, W.~Barnhart, and F.~Forstmayr, ``scikit-rf: An open
  source python package for microwave network creation, analysis, and
  calibration [speaker?s corner],'' {\em IEEE Microwave Magazine}, vol.~23,
  no.~1, pp.~98--105, 2022.

\bibitem{Carter2016}
F.~W. Carter, T.~S. Khaire, V.~Novosad, and C.~L. Chang, ``scraps: An
  open-source python-based analysis package for analyzing and plotting
  superconducting resonator data,'' {\em IEEE Transactions on Applied
  Superconductivity}, vol.~27, pp.~1--5, June 2017.

\bibitem{Pozar:882338}
D.~M. Pozar, {\em {Microwave engineering; 3rd ed.}}
\newblock Hoboken, NJ: Wiley, 2005.

\bibitem{swenson2013}
L.~J. Swenson, P.~K. Day, B.~H. Eom, H.~G. Leduc, N.~Llombart, C.~M. McKenney,
  O.~Noroozian, and J.~Zmuidzinas, ``Operation of a titanium nitride
  superconducting microresonator detector in the nonlinear regime,'' {\em arXiv
  preprint arXiv:1305.4281v1}, 2013.

\bibitem{probst2014}
S.~Probst, F.~B. Song, P.~A. Bushev, A.~V. Ustinov, and M.~Weides, ``Efficient
  and robust analysis of complex scattering data under noise in microwave
  resonators,'' {\em arXiv preprint arXiv:1410.3365v2}, 2014.

\bibitem{dai2023method}
X.~Dai, X.~Liu, Q.~He, Y.~Chen, Z.~Mai, Z.~Shi, W.~Guo, Y.~Wang, M.~R. Vissers,
  and J.~Gao, ``New method for fitting complex resonance curve to study
  nonlinear superconducting resonators,'' {\em Superconductor Science and
  Technology}, vol.~36, no.~1, p.~015003, 2023.

\bibitem{vissers2015}
M.~Vissers, M.~Sandberg, T.~Ohki, J.~Kline, M.~Weides, J.~Gao, and D.~Pappas,
  ``Frequency-tunable superconducting resonators via nonlinear kinetic
  inductance,'' {\em Applied Physics Letters}, vol.~107, no.~6, p.~062601,
  2015.

\bibitem{basu_thakur2019superconducting}
S.~Shu, N.~Klimovich, A.~D. Beyer, R.~B. Thakur, H.~G. Leduc, and P.~Day.,
  ``Nonlinearity and dissipation in superconducting resonators at single-photon
  levels,'' {\em Phys. Rev. Research}, 2021.

\end{thebibliography}
\end{document}